\documentclass[a4]{article}
\title{Fluctuation Scaling in Neural Spike Trains}
\date{ }

\usepackage{amsmath,amssymb}
\usepackage{a4wide}
\usepackage{graphicx}
\usepackage{multirow}
\usepackage{comment}

\begin{document}
\maketitle

% Enter the first author's name and address:
%\centerline{\scshape  Shinsuke Koyama}
\centerline{Shinsuke Koyama}
\medskip
{\footnotesize
% please put the address of the first author
   \centerline{Department of Statistical Modeling, The Institute of Statistical Mathematics}
   \centerline{10-3 Midoricho, Tachikawa, Tokyo, Japan}
   \centerline{
   Department of Statistical Science, SOKENDAI (The Graduate University for Advanced Studies)
   }
    \centerline{10-3 Midoricho, Tachikawa, Tokyo, Japan}
%   \centerline{ERATO Sato Live Bio-Forecasting Project}
%   \centerline{Japan Science and Technology Agency (JST), Kyoto, Japan}
%   \centerline{Advanced Telecommunications Research Institute International, Kyoto, Japan}
} % Do not forget to end the {\footnotesize by the sign }

\medskip

\centerline{Ryota Kobayashi}
\medskip
{\footnotesize
 % please put the address of the second  and third author
   \centerline{Principles of Informatics Research Division, National Institute of Informatics}
   \centerline{2-1-2 Hitotsubashi, Chiyoda-ku, Tokyo, Japan}
  \centerline{Department of Informatics, SOKENDAI (The Graduate University for Advanced Studies)}
   \centerline{2-1-2 Hitotsubashi, Chiyoda-ku, Tokyo, Japan}
}

\bigskip

\begin{abstract}
Fluctuation scaling has been observed universally in a wide variety of phenomena. In time series that describe sequences of events, fluctuation scaling is expressed as power function relationships between the mean and variance of either inter-event intervals or counting statistics, depending on measurement variables. In this article, fluctuation scaling has been formulated for a series of events in which scaling laws in the inter-event intervals and counting statistics were related. We have considered the first-passage time of an Ornstein-Uhlenbeck process and used a conductance-based neuron model with excitatory and inhibitory synaptic inputs to demonstrate the emergence of fluctuation scaling with various exponents, depending on the input regimes and the ratio between excitation and inhibition. Furthermore, we have discussed the possible implication of these results in the context of neural coding.
\end{abstract}

%%%%%%%%%%%%%%%%%%%%%%%%%%%%%%%%%%%%%%%%%%%%%
\section{Introduction}  \label{sec:intro}
%%%%%%%%%%%%%%%%%%%%%%%%%%%%%%%%%%%%%%%%%%%%%
Fluctuation scaling has been observed in a wide range of disciplines.  
It was first observed in ecological systems by Taylor as an empirical power function relationship between the variance and mean of the number of species individuals \cite{Taylor61}. 
Since then, fluctuation scaling has been demonstrated in many other fields, including infectious diseases transmission, cancer metastasis, chromosomal structure, and transportation network traffic \cite{Anderson89,Kendal87,Kendal04,Fronczak10,Barabasi04}, thus demonstrating the univserality of this law. \cite{Eislter08} give a comprehensive review. 

Herein we have considered fluctuation scaling for point processes. A point process is a stochastic process that describes a series of event times $-\infty<t_1<t_2<\cdots<t_n<\infty$, or, in other words, the number of events $N_{(s,t]}$ in a given interval $(s,t]$ \cite{Cox66,Daley02,Snyder75}.
Point processes are used to model a wide variety of phenomena, including neural spike trains, earthquake occurrences, and customer arrivals at a service window \cite{Johnson96,Ogata88,Bremaud81}.

Here we have proposed
fluctuation scaling formulae for a sequence of events, which is expressed as power function relationships between the mean and variance of either the inter-event interval or counting statistics.
For an introduction to the fluctuation scaling law, consider a Poisson process with a rate $\lambda$ for which  
the probability density function of the inter-event interval $x_i:=t_i-t_{i-1}$ is given by the exponential distribution: 
\begin{equation}
f(x) = \lambda e^{-\lambda x},
\label{eq:exponential}
\end{equation}
and the probability distribution of the event count $N_{\Delta}:=N_{(t,t+\Delta]}$ is given by the Poisson distribution:
\begin{equation}
P(N_{\Delta}=n) = \frac{(\lambda\Delta)^{n}}{n!}e^{-\lambda\Delta}.
\label{eq:Poisson}
\end{equation}
The variances in Eqs.~(\ref{eq:exponential}) and (\ref{eq:Poisson}) are given by power functions of the mean as 
$\mathrm{Var}(X)=\mathrm{E}(X)^2$ and $\mathrm{Var}(N_{\Delta})=\mathrm{E}(N_{\Delta})$, respectively.
The fluctuation scaling shown here generalizes these scaling relationships between the mean and variance in both interval and counting statistics using an arbitrary scale factor and exponent. 

In this article, we have focused on the scaling law exponent and have investigated the effect of the underlying mechanism of event occurrences on the exponent.
To address this issue, we have analyzed the first-passage time of an Ornstein-Uhlenbeck (OU) process and a conductance-based neuron model, and have demonstrated the emergence of fluctuation scaling with various exponents under certain conditions. 
Our results suggests that the conventional assumption of proportional relationship between the spike count mean and variance, a fundamental fact of neural coding \cite{Averbeck09}, could lead to the wrong conclusion regarding the variability of neural responses.

%%%%%%%%%%%%%%%%%%%%%%%%%%%%%%%%%%%%%%%%%%%%%%
\section{Fluctuation scaling}
\label{sec:formalism}
%%%%%%%%%%%%%%%%%%%%%%%%%%%%%%%%%%%%%%%%%%%%%%

Consider a sequence of events in which the inter-event intervals are independent and identically distributed with a  mean $\mu$ and variance $\sigma^2$. 
Fluctuation scaling in this interval statistics is described by the following power function relationship between $\mu$ and $\sigma^2$:
\begin{equation}
\sigma^2 = \phi\mu^{\alpha},
\label{eq:scaling-interval}
\end{equation}
where $\phi$ is the scale factor that controls the overall amplitude of the variance and $\alpha$ is the exponent that controls how the variance is scaled by the mean. 
For $\alpha=2$, the scale factor $\phi$ corresponds to the squared coefficient of variation, for which the value is unity in a Poisson process. 
In contrast, $\alpha>2(<2)$ implies a tendency for the event occurrence timing to be over (under) dispersed for large means and under (over) dispersed for small means.

Next, consider the counting statistics. 
Let $N_{\Delta}$ denote the number of events in a counting window of duration $\Delta$. 
For a large counting window relative to the mean inter-event interval $\Delta\gg\mu$, 
the mean and variance of $N_{\Delta}$ asymptotically become $\Delta/\mu$ and $\sigma^2\Delta/\mu^3$, respectively \cite{Cox62}. 
Accordingly, if the interval statistics obeys the scaling law (\ref{eq:scaling-interval}), for a large $\Delta/\mu$ the variance of $N_{\Delta}$ will asymptotically exhibit the scaling law:
\begin{equation}
\mathrm{Var}(N_{\Delta}) \sim \phi\Delta^{1-\beta}\mathrm{E}(N_{\Delta})^{\beta},
\label{eq:scaling-count}
\end{equation}
where the exponent $\beta$ correlates with that of the interval statistics via the scaling relationship:
\begin{equation}
\beta=3-\alpha.
\label{eq:scaling-exponent}
\end{equation}
In the counting statistics, the linear relationship between the mean and variance is maintained only if $\alpha=2$. 
The relationship between the mean and variance is sublinear if $\alpha>2$ and superlinear if $\alpha<2$. 

The scaling law in the counting statistics is obtained for a sufficiently large window relative to the mean inter-event interval,  $\Delta \gg \mu$. 
In the numerical studies presented in section~\ref{sec:statanalysis}, however, 
we found that an average of five events falling in a counting window is enough for Eq.~(\ref{eq:scaling-count}) to apply.

We should emphasize that the scaling law (\ref{eq:scaling-count}) is obtained when the mean inter-event interval is changed and $\Delta$ is fixed. It is also possible to have another scaling law. For instance, we can have a simple linear relationship between the count mean and variance for stationary renewal processes when $\Delta$ is changed \cite{Cox62}. In this article, we consider the scaling law (\ref{eq:scaling-count}), because the count mean is modulated and $\Delta$ is fixed in the analysis of nonstationary event sequences, which is discussed in section~\ref{sec:statanalysis}.

%%%%%%%%%%%%%%%%%%%%%%%%%%%%%%%%%%%%%%%%%%%%%
\section{First-passage time analysis}
%%%%%%%%%%%%%%%%%%%%%%%%%%%%%%%%%%%%%%%%%%%%%

In this section, we analyze the first-passage time to a threshold using an OU process and a conductance-based neuron model to investigate under which conditions fluctuation scaling (\ref{eq:scaling-interval}) emerges. 

\subsection{OU process}
We consider an OU process described by the following stochastic differential equation \cite{Kampen92}:
\begin{equation}
\frac{dV(t)}{dt} = -\frac{V(t)}{\tau} + a + b\xi(t), \quad V(0)=v_r,
\label{eq:oup_org}
\end{equation}
where $\xi(t)$ is Gaussian white noise with $\mathrm{E}[\xi(t)]=0$ and $\mathrm{E}[\xi(t)\xi(t')]=\delta(t-t')$.
If $V(t)$ exceeds a threshold $\theta>0$, an event occurs and $V(t)$ is immediately reset to $v_r$. 
By rescaling 
$(V-v_r)/(\theta-v_r) \to V$ and $t/\tau \to t$, the model parameters are rescaled as 
$(a\tau-v_r)/(\theta-v_r) \to a$, $b\sqrt{\tau}/(\theta-v_r) \to b$, $\theta \to 1$ and $v_r\to0$.
Accordingly, Eq.~(\ref{eq:oup_org}) is rewritten as follows:
\begin{equation}
\frac{dV(t)}{dt} = -V(t) + a + b\xi(t), \quad 
V(0)= 0,
\label{eq:oup}
\end{equation}
which has two free parameters: $(a,b)$.
We can analyze Eq.~(\ref{eq:oup}) without loss of generality.

The stochastic integration of Eq.~(\ref{eq:oup}) without the threshold condition yields the solution of $V(t)$:
\begin{equation}
V(t) = a(1-e^{-t})+b\int_0^te^{s-t}\xi(s)ds,
\end{equation}
from which the mean and variance of $V(t)$ are respectively obtained as follows:
\begin{equation}
\mathrm{E}[V(t)] = a(1-e^{-t}),
\label{eq:meanv}
\end{equation}
and
\begin{equation}
\mathrm{Var}[V(t)] = \frac{b^2}{2}(1-e^{-2t}).
\end{equation}
Depending on the values of asymptotic mean $a$ and fluctuation $b$ relative to the threshold, the following three asymptotic regimes are considered (a similar regime division can found in \cite{Wan82,Ditlevsen05}):
\begin{enumerate}
\item[a)]
Suprathreshold regime ($a \gg 1$) with small fluctuations ($b\ll1$), in which the threshold is exceeded mainly because of drift $a$.
\item[b)]
Subthreshold regime ($1-a\gg b$) with small fluctuations ($b\ll 1$), in which the threshold is relatively rarely passed because of small fluctuations in $V(t)$. 
\item[c)]
Threshold regime ($a \sim 1$) with large fluctuations $(b\gg1)$, in which the threshold is strongly exceeded because of the large fluctuations in $V(t)$.
\end{enumerate}
The first-passage time analysis of the OU process in the three asymptotic regimes is described as follows. The results are summarized in Table~\ref{tbl:analytical}.

\begin{table}
\begin{center}
\begin{tabular}{rlcccc}
\hline\noalign{\smallskip}
   & Regime & \multicolumn{2}{c}{Condition} & $\alpha$ & $\phi$ \\
\noalign{\smallskip}\hline\noalign{\smallskip}
\multirow{ 2}{*}{a.} & \multirow{2}{*}{Suprathreshold} & \multirow{ 2}{*}{$a\gg1,b\ll1$} & $b$ : fixed & 3 & $b^2$ \\
  &   &  &  $b=c\sqrt{a}$ & 2 & $c^2$  \\
b. & Subthreshold & \multicolumn{2}{c}{$1-a\gg b, b\ll 1$} & 2 & 1 \\ 
c. & Threshold      &  \multicolumn{2}{c}{$a\sim 1, b\gg 1$} & 1 & $2\log 2$ \\
\noalign{\smallskip}\hline
\end{tabular}
\end{center}
\caption{Various scaling exponents $\alpha$ and factors $\phi$ emerged in the first-passage time of the OU process.}
\label{tbl:analytical} 
\end{table}

%%%%%%%%%%%%%%%%%
\subsubsection{Suprathreshold regime} \label{sec:supra}

For $b\ll1$ and $a-1\gg b$, the mean and variance of the first-passage time have been evaluated in \cite{Wan82} as follows:
\begin{equation}
\mu \sim \log\frac{a}{a-1} - \frac{b^2}{4}\Bigg[\frac{1}{(a-1)^2}-\frac{1}{a^2}\Bigg],
\label{eq:mean_supra}
\end{equation}
and 
\begin{equation}
\sigma^2 \sim \frac{b^2}{2}\Bigg[\frac{1}{(a-1)^2}-\frac{1}{a^2}\Bigg].
\label{eq:var_supra}
\end{equation}
A further assumption of $a\gg1$ and expanding Eqs.~(\ref{eq:mean_supra}) and (\ref{eq:var_supra}) with respect to $1/a$ and while selecting the leading terms yields
\begin{equation}
\mu \sim \frac{1}{a}, \quad \sigma^2 \sim \frac{b^2}{a^3}.
\label{eq:mean_var_supra}
\end{equation}
Thus, the variance of the first-passage time obeys the scaling law (\ref{eq:scaling-interval}) with the exponent $\alpha=3$ and the factor $\phi=b^2$, if the mean is modulated by changing $a$ while keeping $b$ unchanged. 

This scaling law may also be obtained as follows. 
For $a\gg1$ and $b\ll 1$, $|V(t)|\ll a$ and 
Eq.~(\ref{eq:oup}) is approximated to Brownian motion with the drift:
\begin{eqnarray}
\frac{dV(t)}{dt} = a\bigg[-\frac{V(t)}{a}+1\bigg] + b\xi(t) 
\approx a + b\xi(t),
\end{eqnarray}
for which the first-passage time probability distribution can be obtained analytically as the inverse Gaussian distribution \cite{Tuckwell88}. The density function is given by  
\begin{equation}
f(x;a,b) = \frac{1}{\sqrt{2\pi b^2x^3}}\exp\bigg[-\frac{(1-ax)^2}{2b^2x}\bigg],
\end{equation}
the mean and variance of which correspond to Eq.~(\ref{eq:mean_var_supra}).

We  can consider another situation in which both $a$ and $b$ are changed. 
A typical situation is that $b$ is modulated by $a$ in a square root manner, $b = c\sqrt{a}$, $c$ being a constant (which is realized by diffusion approximation of Poisson inputs).
Substituting it into Eq.~(\ref{eq:mean_var_supra}) yields the scaling law:
\begin{equation}
\sigma^2 \sim c^2\mu^2,
\end{equation}
whose exponent ($\alpha=2$) differs from that obtained by keeping $b$ unchanged.

%%%%%%%%%%%%%%%%%%%%%%
\subsubsection{Subthreshold regime}

For $1-a\gg b$ and $b\ll1$, the asymptotic mean and variance of the first-passage time were derived in \cite{Wan82} as follows: 
\begin{equation}
\mu \sim \frac{b\sqrt{\pi}}{1-a}\exp\bigg[\frac{(1-a)^2}{b^2}\bigg],
\label{eq:mean_exp}
\end{equation}
and 
\begin{equation}
\sigma^2 \sim \frac{b^2\pi}{(1-a)^2}\exp\bigg[\frac{2(1-a)^2}{b^2}\bigg],
\label{eq:var_exp}
\end{equation}
which follow the fluctuation scaling law (\ref{eq:scaling-interval}) with $\alpha=2$ and $\phi=1$.
This scaling law does not depend on the way in which $a$ and $b$ are changed. 
In fact, in this limit, the first-passage time probability distribution asymptotically becomes an exponential distribution with the mean (\ref{eq:mean_exp}) \cite{Nobile85,Ricciardi88}, such that the first-passage time sequence becomes a Poisson process.

%%%%%%%%%%%%%%%%%%%%%%%%%%%%%%%
\subsubsection{Threshold regime}

The Laplace transformation of the probability density function $f(x;a,b)$ for the OU process first-passage time has analytically been derived in \cite{Tuckwell88,Siegert51,Roy69} as follows:
\begin{eqnarray}
G(s) = \int_0^{\infty}e^{-sx}f(x;a,b)dx  
=
\frac{\Psi\big(\frac{s}{2},\frac{1}{2},(\frac{a}{b})^2)}{\Psi\big(\frac{s}{2},\frac{1}{2},(\frac{1-a}{b})^2)},
\label{eq:laplace}
\end{eqnarray}
where $\Psi(x,y,z)$ is a confluent hypergeometric function of the second kind \cite{Abramowitz65}. 
For $z\ll1$, $\Psi(x,\frac{1}{2},z)$ is evaluated as follows:
\begin{equation}
\Psi(x,\frac{1}{2},z) \sim \frac{\sqrt{\pi}}{\Gamma(x+\frac{1}{2})}
- \frac{2\sqrt{\pi}}{\Gamma(x)}z^{\frac{1}{2}}.
\label{eq:hypergeometric}
\end{equation}
Using Eqs.~(\ref{eq:laplace}) and (\ref{eq:hypergeometric}), the mean and variance of the first-passage time for $a\sim1$ and $b\gg 1$ are obtained as follows:
\begin{equation}
\mu = -\lim_{s\to0}\frac{dG(s)}{ds} \sim \frac{\sqrt{\pi}}{b},
\end{equation}
and
\begin{eqnarray}
\sigma^2 
=
\lim_{s\to0}\frac{d^2G(s)}{ds^2} - \mu^2 
\sim \frac{2\sqrt{\pi}\log2}{b}.
\end{eqnarray}
Accordingly, in this limit, the fluctuation scaling law (\ref{eq:scaling-interval}) emerges with  $\alpha=1$ and $\phi=2\log2$.
Note that $a$ does not appear in the leading terms of $\mu$ and $\sigma^2$, suggesting that the mean and variance of the first-passage time are modulated mainly by changing $b$ in this regime.

%%%%%%%%%%%%%%%%%%%%%%%%%%%%%%%%%%%%
\subsubsection{Numerical results}

\begin{figure}[tb]
\begin{center}
\includegraphics[width=10cm]{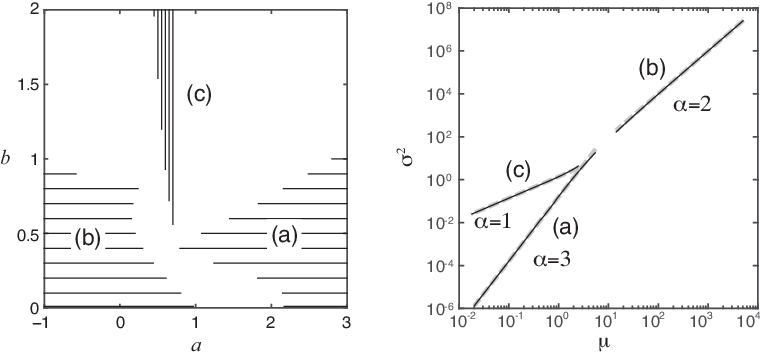}
\end{center}
\caption{
Left: the parameter space $(a,b)$. The mean $\mu$ and variance $\sigma^2$ of the OU process first-passage time were computed by changing $a$ and $b$ along solid lines 
in (a) suprathreshold, (b) subthreshold and (c) threshold regimes.  
Right: Log-log plot of the variance $\sigma^2$ against the mean $\mu$. 
Solid lines represent the exact mean-variance relationship mapped from the left panel. 
Gray dashed lines represent the scaling laws obtained through an asymptotic analysis (summarized in Table~\ref{tbl:analytical}), which exhibit good agreement with the exact mean-variance relationships. 
}
\label{fig:loglog}
\end{figure}

The scaling laws summarized in Table~\ref{tbl:analytical} were obtained using the three asymptotic regimes. 
To examine the extent to which these scaling laws capture the actual mean-variance relationships, we have compared it with the exact variance of the OU process first-passage time as computed using series expansion formulas \cite{Ricciardi88,Keilson75,Inoue95}.
Figure~\ref{fig:loglog} shows the result:
solid lines in the left panel represent how the set of parameters was changed in $(a,b)$ space, while the right panel plots the exact mean-variance relationships mapped from the left panel (solid lines). 
(Note that there is a unique mapping between $(a,b)$ and $(\mu,\sigma^2)$ \cite{Vilela09}.) 
It is seen that wide areas in $(a,b)$ space are approximately mapped onto the scaling laws obtained in the asymptotic analysis (thick dashed lines), suggesting that the scaling laws provide good descriptions of the mean-variance relationships in the three regimes.

%%%%%%%%%%%%%%%%%%%%%%%%%%%
\subsection{Neuron model}
%%%%%%%%%%%%%%%%%%%%%%%%%%%%

Here we consider a particular interpretation of the first-passage time in terms of neural spike trains. 
In the following subsections, we will describe a neuron model with a realistic synaptic input and a simplified model. 

\subsubsection{A realistic description of synaptic input} 

The membrane potential dynamics at the soma in a model neuron, $V(t)$, is described as follows: 
\begin{equation}
	C\frac{dV(t)}{dt} = -g_L (V(t)-E_L)+ I_{\scalebox{0.7}{AMPA} }+  I_{ \scalebox{0.7}{GABA} }, 
	\label{eq:lif}
\end{equation}
where, $C= 1$ $\mu$F/cm$^2$ is the membrane capacitance, $g_L= 4.52 \times 10^{-2}$ mS/cm$^2$ is the leak conductance, $E_L= -70$ mV is the reversal potential for the leak current, and $I_{\scalebox{0.7}{AMPA} (\scalebox{0.7}{GABA})}$ is the AMPA (GABA) synaptic current. The model neuron generates a spike when the potential $V(t)$ exceeds the spike threshold $\theta$, at which point $V(t)$ is instantaneously reset to $v_r$. 
The synaptic current is described by the conductance input from pre-synaptic neurons as follows: 
\begin{equation}
	\begin{split}
		I_{ \scalebox{0.7}{AMPA} } &= - \sum_{k=1}^{N_E }g_{\scalebox{0.7}{AMPA} } s_{ \scalebox{0.7}{AMPA}, k}(t) (V-V_E), \\		
		I_{ \scalebox{0.7}{GABA} } &= - \sum_{k=1}^{N_I } g_{\scalebox{0.7}{GABA} } s_{ \scalebox{0.7}{GABA}, k}(t) (V-V_I), 	
	\end{split}
	\label{eq:synapse}
\end{equation}
where $N_E$ and $N_I$ are the numbers of excitatory and inhibitory synapses with their respective reversal potentials $V_E$ and $V_I$, $g_{x}$ is the maximal synaptic conductance, $s_{x, k}$ are the gating variables of the $k$-th synapse, and $x$ represents the synaptic component (AMPA or GABA). 
The gating variable of the $x$-synaptic component $s_x$ is described by the first-order kinetics as follows \cite{Destexhe1998}: 
\begin{equation}
	\frac{ds_x}{dt} = \alpha_x [T](t) (1-s_x) - \beta_x s_x, 
	\label{eq:gate}
\end{equation}
where $[T](t)$ is the transmitter concentration in a neuronal cleft, and $\alpha_x$ and $\beta_x$ are the activation and inactivation rates, respectively. 
When the pre-synaptic neuron generates a spike, transmitter accumulates in the cleft such that $[T]= 1$ mM for 1 ms: $[T]$ is subsequently set to 0 before the next spike occurs. 
The spike trains of the excitatory (E) and inhibitory (I) presynaptic neurons were generated by the Poisson process with constant rate $r_{E, I}= \lambda_{E, I}/ N_{E, I}$, where $\lambda_{E, I}$ is the total input rate from the excitatory (E) and inhibitory (I) neurons.
The synaptic parameters were $g_{ \scalebox{0.7}{AMPA} }= 1.2$ nS, $\alpha_{\scalebox{0.7}{AMPA} }= 1.1 \times 10^6$ M$^{-1}$ s$^{-1}$, $\beta_{\scalebox{0.7}{AMPA} }= 670$ s$^{-1}$ for the AMPA synapses; 
and $g_{ \scalebox{0.7}{GABA} }= 0.6$ nS, $\alpha_{\scalebox{0.7}{GABA} }= 5.0 \times 10^6$ M$^{-1}$ s$^{-1}$, $\beta_{\scalebox{0.7}{GABA} }= 180$ s$^{-1}$, for the GABA synapses unless stated. 
The other parameters were $N_E= 2,000$, $N_I= 2,000$, $V_E= 0$ mV, and $V_I= -75$ mV.

%%%%%%%%%%%%%%%%%%%%%%%%%%%%%
\subsubsection{Diffusion approximation}

The neuron model dynamics (\ref{eq:lif}) using the realistic synaptic model (\ref{eq:synapse}) and (\ref{eq:gate}) can be approximated as follows (See Appendix~\ref{sec:reduction} for the derivation):
\begin{equation}
	C\frac{dV(t)}{dt} = -g_{tot}(V-E_{tot}) + \sigma_0\xi(t),
\label{eq:lifgwn}
\end{equation}
where $\xi(t)$ is Gaussian white noise, and 
\begin{equation}
	g_{tot} = g_L + A_{\scalebox{0.7}{AMPA} } \lambda_E + A_{\scalebox{0.7}{GABA} }\lambda_I,
\end{equation}
\begin{equation}
	E_{tot} = (g_L E_L +A_{\scalebox{0.7}{GABA} }\lambda_I V_I)/g_{tot},
\end{equation}
\begin{equation}
	\sigma_0^2 = A_{\scalebox{0.7}{AMPA} }^2\lambda_E E_{tot}^2 + A_{\scalebox{0.7}{GABA} }^2 \lambda_I(E_{tot}-V_I)^2,
\end{equation}
where $\lambda_{E (I)}$ is the total firing rate of the excitatory (inhibitory) pre-synaptic neurons and $A_{\scalebox{0.7}{AMPA} (\scalebox{0.7}{GABA}) }$ represents the effect of a pre-synaptic spike on the AMPA (GABA) input.
By rescaling the membrane potential and the time as 
$(V-v_r)/(\theta-v_r)\to V$ and $g_{tot}t/C\to t$, Eq.~(\ref{eq:lifgwn}) is rescaled as in Eq.~(\ref{eq:oup}), as follows:
\begin{equation}
a = \frac{E_{tot}-v_r}{\theta-v_r},
\label{eq:drift}
\end{equation}
\begin{equation}
b = \frac{\sigma_0}{\sqrt{g_{tot}C}(\theta-v_r)}.
\label{eq:noise}
\end{equation}

%%%%%%%%%%%%%%%%%%%%%%%%%%%%%%%%%%%%%%
\subsubsection{Numerical results}
%%%%%%%%%%%%%%%%%%%%%%%%%%%%%%%%%%%%%%

\begin{figure}[tb]
\begin{center}
\includegraphics[width=10cm]{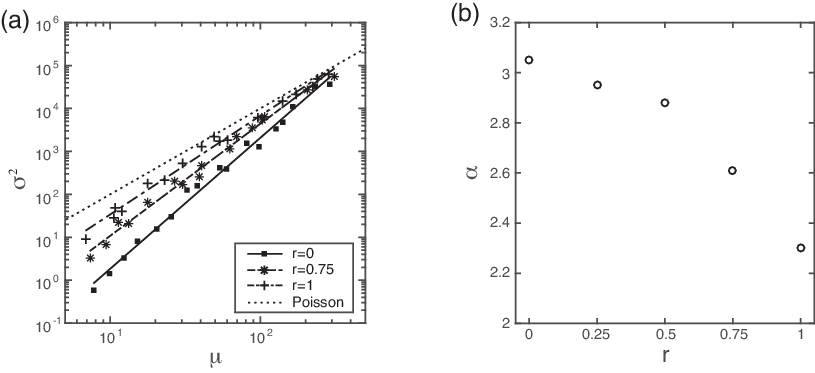}
\end{center}
\caption{The mean-variance relationship for a neuron model with a realistic synaptic input (AMPA and GABA).
(a) Variance as a function of the means of ISIs at different EI ratios, $r$. 
The dotted line represents $\sigma^2=\mu^2$ (i.e., the Poisson case). 
(b) The exponent $\alpha$ as a function of $r$.
}
\label{fig:AMPA}
\end{figure}

We have also considered balanced excitatory and inhibitory synaptic inputs in which both excitatory and inhibitory input rates increased while maintaining a constant ratio $r>0$:
\begin{equation}
\lambda_I = r\lambda_E.
\end{equation}
We simulated spike trains using the model (\ref{eq:lif}), (\ref{eq:synapse}) and (\ref{eq:gate}), and computed the means and variances of inter-spike intervals (ISIs) for different $\lambda_E$. 
Figure~\ref{fig:AMPA}a plots the variances of these ISIs against the means for different values of $r$.
We observed that the means and variances exhibited an approximate linear relationship on a log-log scale. 
We performed a linear regression analysis of the $\log\mathrm{Var(ISI)}$ on $\log\mathrm{E(ISI)}$. 
The fitted slope (i.e., the exponent $\alpha$) is plotted as a function of $r$ in Figure~\ref{fig:AMPA}b. 
We observed that this exponent was $\alpha\approx3$ when $r=0$ (i.e., excitation is dominant). 
The exponent decreased toward $\alpha\approx 2$ as the inhibition increased. 

\begin{figure}
\begin{center}
\includegraphics[width=10cm]{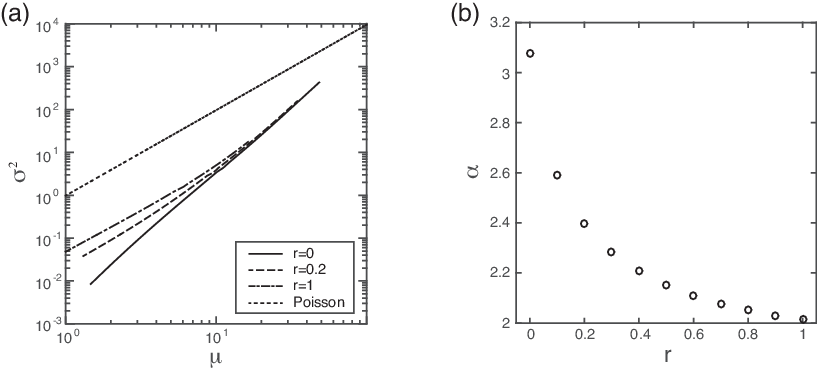}
\end{center}
\caption{
The mean-variance relationship for a neuron model with white noise input. 
(a) Variance as a function of the means of ISIs at different EI ratios, $r$. 
The dotted line represents $\sigma^2=\mu^2$ (i.e., the Poisson case).
(b) The exponent $\alpha$ as a function of $r$. The exponent $\alpha$ is obtained through linear regression of $\log\sigma^2$ on $\log\mu$.  
}
\label{fig:balance-oup}
\end{figure}

We computed the means and variances of the ISIs using series expansion formulas \cite{Ricciardi88,Keilson75,Inoue95} in the simplified model given by (\ref{eq:oup}), (\ref{eq:drift}) and (\ref{eq:noise}), and obtained the similar results as those achieved using 
the model with realistic synaptic inputs (Figure~\ref{fig:balance-oup}). 
Although the exponent $\alpha$ varies in similar ranges in both models, 
the dependences of $\alpha$ on $r$ are different: it is a curve with negative curvature for the model with realistic synaptic inputs (Figure~\ref{fig:AMPA}b), while it is a curve with positive curvature for the simplified model (Figure~\ref{fig:balance-oup}b). 
This difference might be due to neglecting the synaptic time constant, which is the main assumption for deriving the simplified model (Appendix~\ref{sec:reduction}).

%%%%%%%%%%%%%%%%%%%%%%%%%%%%%%%%%%%%%%%%%%%%%%%
\section{Impact of fluctuation scaling on the statistical analysis of neural data}
\label{sec:statanalysis}
%%%%%%%%%%%%%%%%%%%%%%%%%%%%%%%%%%%%%%%%%%%%%%%

Both the analyses of the neuron model and of the OU process first-passage time revealed that exponent $\alpha$ from the interval statistics can hold different values depending on  the input regimes or on the ratio between the excitatory and inhibitory synaptic inputs. 
A consequence from these results is that the variance of spike count can exhibit ``nontrivial" scaling laws (\ref{eq:scaling-count})-(\ref{eq:scaling-exponent}) that depart from simple linear relationships. 
In this section, we will examine the extent to which the scaling properties of neural responses affect the neural data analysis.
In particular, we will demonstrate that the conventional assumption of linear relationship between the spike count mean and variance could lead to the wrong conclusion regarding the variability of neural responses.

Recent experimental data analysis suggested that apparent variability in the observed spike trains can be attributed to two sources:
spiking variability, which effectively acts as measurement noise and cross-trial fluctuations in firing rates, which correlate with behavior or perception \cite{ChurchlandMM10,ChurchlandAK11,ChurchlandMM12}. 
Churchland et al. (2011) proposed a method for segregating response variability into spiking variability and firing rate variability, according to the law of total variance for doubly stochastic processes  \cite{ChurchlandAK11}. 
Here we will critically analyze their method and demonstrate how their method could lead to a wrong conclusion as a result of their assumption that the spike count variance is proportional to the mean. 

Here $N_{\Delta}(t)$ is set as the number of spikes in the counting window of duration $\Delta$ centered at time $t$ and $\lambda(t)$ is the mean firing rate in this window. 
We will assume that $\lambda(t)$ is also a random variable, thus allowing a different realized $\lambda(t)$ in each trial. 
According to the law of total variance, the total variance of $N_{\Delta}(t)$ can be decomposed into two components:
\begin{eqnarray}
\mathrm{Var}(N_{\Delta}(t)) 
&=& \mathrm{Var}(\lambda(t)\Delta) +  \mathrm{E}[\mathrm{Var}(N_{\Delta}(t)|\lambda(t))].
\label{eq:totalvar}
\end{eqnarray}
The first term on the right side of Eq.~(\ref{eq:totalvar}) represents the cross-trial variability of the firing rate, whereas the last term in Eq.~(\ref{eq:totalvar}) represents the spiking variability. 
In accordance with \cite{ChurchlandAK11}, we refer to the former as the ``variance of the conditional expectation"  (VarCE) and the latter as the ``point process variance" (PPV).
If the firing rate is the same in each trial, then the VarCE is zero and the total spike count variance is attributed solely to the PPV. 
If the firing rate differs in each trial, the VarCE will capture this variance. 
To obtain an estimate of the VarCE from neural data, we can calculate the sample spike count variance and subtract the estimate of the PPV. 
To obtain this estimated PPV, it was assumed in \cite{ChurchlandAK11} that the spike count variance is proportional to the mean:
\begin{equation}
\mathrm{Var}(N_{\Delta}(t)|\lambda(t)) = \phi \mathrm{E}(N_{\Delta}(t)|\lambda(t)).
\label{eq:prop}
\end{equation}
From Eqs.~(\ref{eq:totalvar}) and (\ref{eq:prop}), the VarCE then becomes 
\begin{equation}
\mathrm{Var}(\lambda(t)\Delta) = \mathrm{Var}(N_{\Delta}(t)) - \phi \mathrm{E}(N_{\Delta}(t)), 
\label{eq:varce2}
\end{equation}
where both terms on the right side of the above equation are easily estimated using the sample mean $\bar{N}_{\Delta}(t)$ and variance, $s^2_{N_{\Delta}}(t)$, of the cross-trial spike counts.  
If we know $\phi$, then the estimated VarCE, $s^2_{\lambda}(t)$, is
\begin{equation}
s^2_{\lambda}(t) = s^2_{N_{\Delta}}(t) - \phi \bar{N}_{\Delta}(t).
\label{eq:varce_est}
\end{equation}
As was done in \cite{ChurchlandAK11}, we found the time window with the smallest variance to mean ratio (i.e., the Fano factor), and took the Fano factor from this epoch as an estimate of $\phi$, which ensures a positive estimated VarCE throughout the trial. (See \cite{ChurchlandAK11} in more detail.) 

\begin{figure}[tb]
\begin{center}
\includegraphics[width=10cm]{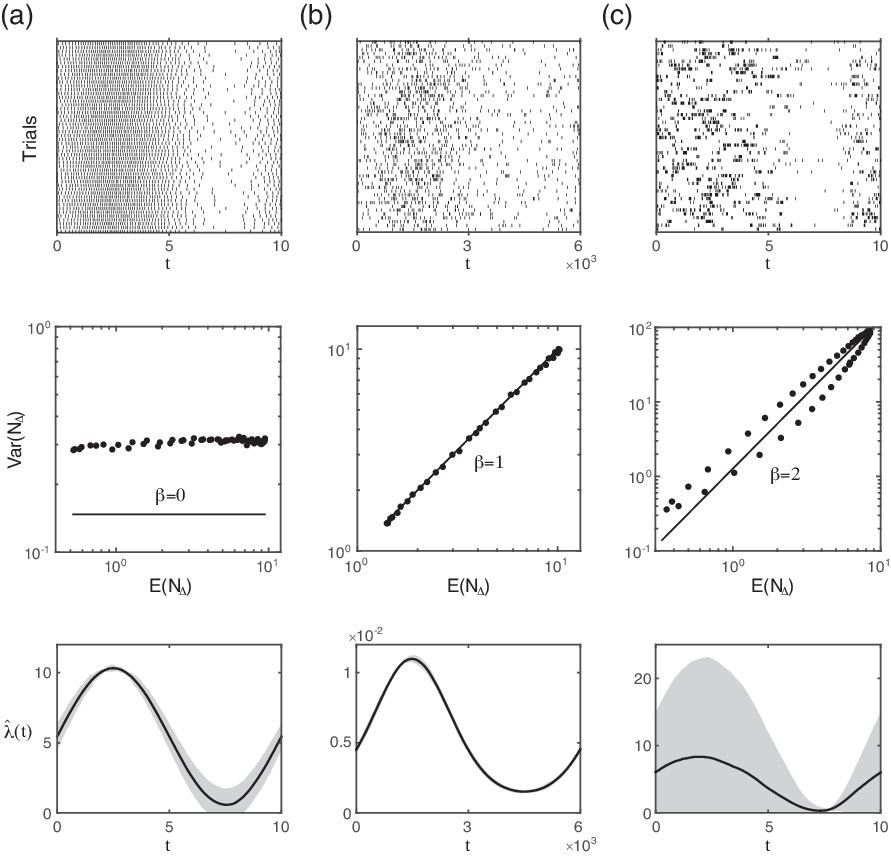}
\end{center}
\caption{
Results of the OU process first-passage time to a threshold in the suprathreshold (a), subthreshold (b), and threshold regimes (c). 
The mean spike count, $\mathrm{E}(N_{\Delta})$, was temporally modulated by the variations of $a$ and $b$ (\ref{eq:periodic}).
Top: raster plots of 50 spike trains. 
Middle: the mean-variance relationship using the counting statistics. 
Filled circles represent the sample means and variances computed with $10^4$ trials. 
Solid lines represent the theoretical scaling law (\ref{eq:scaling-count}).
Bottom: solid lines represent the estimated firing rate $\hat{\lambda}(t)$ and gray regions represent $\pm 2 \sqrt{s^2_{\lambda}(t)}$; $s^2_{\lambda}(t)$ is the estimated VarCE (\ref{eq:varce_est}).
}
\label{fig:oup-nonstationary}
\end{figure}

We demonstrated using simulated spike trains that the estimator (\ref{eq:varce_est}) might fail to capture the actual VarCE. 
In our numerical study, the OU process (\ref{eq:oup}) was simulated in the three regimes using time-varying parameters given by
\begin{eqnarray}
\begin{array}{rll}
\mathrm{Suprathreshold}: & a=6+5\sin\frac{2\pi}{10}t, & b=0.4 \\
\mathrm{Subthreshold}: & a=0.1+0.1\sin\frac{2\pi}{6000}t, & b=0.4  \\
\mathrm{Threshold}: & a=0.6, & b=10+9\sin\frac{2\pi}{10}t   
\end{array}
\label{eq:periodic}
\end{eqnarray}
For each regime, $10^4$ spike trains were numerically generated. 
The raster plots of 50 spike trains are displayed in Figure~\ref{fig:oup-nonstationary} (top). 
Note that the same parameters were used in each trial, and thus the firing rates are identical in each trial, such that the theoretical value of VarCE was zero. 

The spike count mean and variance were computed using the $10^4$ spike trains and a sliding window whose length, $\Delta$, was chosen so that an average of five spikes fell within the window per trial. 
The firing rate did not change drastically in each window, which allowed us to apply Eq.~(\ref{eq:scaling-count}).  
Figure~\ref{fig:oup-nonstationary} (middle) also plots the variance against the mean on a log-log scale (filled circles) and shows that the variance was well described using the theoretical scaling relationship  (\ref{eq:scaling-count}) (lines). 
The variance was proportional to the mean ($\beta=1$) in the subthreshold regime. 
The mean-variance relationship was sublinear ($\beta=0$) in the suprathrehold regime, whereas it was superlinear ($\beta=2$) in the threshold regime. 
The estimated firing rate, $\hat{\lambda}(t) = \bar{N}_{\Delta}(t)/\Delta$, is displayed together with the estimated VarCE (\ref{eq:varce_est}) in Figure~\ref{fig:oup-nonstationary} (bottom; solid lines and gray regions represent $\hat{\lambda}(t) \pm 2\sqrt{s^2_{\lambda}(t)}$, respectively). 
The estimated VarCE in the subthreshold regime is shown to be near zero (b), although it significantly departed from zero in the suparthrehold (a) and threshold (c) regimes. 
The estimated VarCE increased as the firing rate decreased in the suprathreshold regime, but it increased as the firing rate increased in the threshold regime. 
Note that the actual value of VarCE for all the three cases was zero; 
the finite estimated VarCE values resulted from a wrong assumption (\ref{eq:prop}), although the actual relationship between the spike count mean and variance was not linear in the suprathreshold and threshold regimes.

%%%%%%%%%%%%%%%%%%%%%%%%%%%%%%%%%%%%%%%%%%%%%%%
\section{Discussion}
%%%%%%%%%%%%%%%%%%%%%%%%%%%%%%%%%%%%%%%%%%%%%%%

This article describes the formulation of fluctuation scaling for sequences of events. 
This fluctuation scaling is expressed as power function relationships between the means and variances in either the interval statistics (\ref{eq:scaling-interval}) or counting statistics (\ref{eq:scaling-count}), which are linked via the scaling relationship (\ref{eq:scaling-exponent}).
Furthermore, this article demonstrates that the first-passage time to a threshold exhibits fluctuation scaling in which the exponent depends on the OU process input regimes.
In the suprathreshold regime, in which threshold crossing is mainly caused by positive drift, the event occurrence tends to be regular, resulting in the exponent $\alpha=3$, whereas in the subthreshold regime with small fluctuations, threshold crossing is relatively rare, and the event sequences exhibit Poisson statistics with $\alpha=2$.
In the threshold regime, in which threshold passing is largely induced by large OU process fluctuations, the first-passage time is more variable, resulting in $\alpha=1$.
We also examined fluctuation scaling in a conductance-based neuron model with balanced excitatory and inhibitory synaptic inputs and showed that the excitation to inhibition ratio modulates the scaling exponent; in particular, when excitation is dominant, the exponent becomes $\alpha\approx3$ and decreases toward $\alpha\approx2$ as inhibition increases. 

We note that many of the mathematical results concerning the issue of OU process first-passage times were derived long ago (\cite{Burkitt06} and references therein).
Our OU process results mostly rely on these earlier findings. 
However, to our best knowledge, no previous reports have addressed this problem systematically from the viewpoint of fluctuation scaling (\ref{eq:scaling-interval}), particularly in relation to the exponent $\alpha$.
We therefore believe that this article presents a novel viewpoint on the first-passage time problem. 

An important implication of our results is that renewal processes do not necessarily imply a proportional relationship between the event count mean and variance (proportionality is maintained only if $\alpha=2$). 
In the field of neural coding, a proportional relationship between the spike count mean and variance is considered a fundamental fact that is relevant almost anywhere in the brain \cite{Averbeck09}. 
Our analysis of a neuron model as well as the OU process first-passage time revealed that the interval statistics exponent is not necessarily $\alpha=2$; this means that the spike count statistics can significantly deviate from the proportional relationship, and therefore analysis methods based on this assumption can fail (see section~\ref{sec:statanalysis}). 

One possible application of fluctuation scaling may be characterization of the ``intrinsic" variability of neuronal firing. 
Troy and Robson found that in {\it in vivo} recordings, steady discharges of X retinal ganglion cells in response to stationary visual patterns exhibited the scaling law in interval statistics \cite{Troy92}, for which the exponent was $\alpha\approx 3$ in our formulation. 
In contrast, cortical spike trains exhibit an approximately proportional relationship between the spike count variance and the mean \cite{Tolhurst81,Shadlen98}, suggesting that $\alpha\approx 2$. 
One might speculate that a difference in the scaling exponent reflects the electrophysiological properties of individual cells or their networks. 
Further investigations are needed to clarify the relationship between the scaling exponent and neurophysiological properties. 
Another theoretical question is whether fluctuation scaling with various exponents emerges from the dynamics of a network of spiking neurons. 
In the future work, it would be interesting to investigate how network fluctuations are translated into fluctuation scaling in a self-consistent framework \cite{Lerchner06}.

%For acknowledgements section, please don't number the section, please begin it with \section*{Acknowledgements}
\section*{Acknowledgments} 
This work was emerged from the discussion during the neural coding workshop in Versailles. 
We would like to appreciate Philippe Lucas, Jean-Pierre Rospars, Petr Lansky, Chris Christolodou, Lubomir Kostal, and all the staffs who worked for the workshop. 
S.K. would like to thank Prof. Jianfeng Feng for valuable comments on the draft.

%%%%%%%%%%%%%%%%%%%%%%%%%%%%%%%%%%%%%%%%%%%%%%%%%%%%
\appendix
\section{Diffusion approximation for a neuron model with a realistic synaptic input} \label{sec:reduction}
%%%%%%%%%%%%%%%%%%%%%%%%%%%%%%%%%%%%%%%%%%%%%%%%%%%%
In the first-order kinetic model (\ref{eq:gate}), the conductance change after a presynaptic input can be written as follows:
\begin{eqnarray}
	\begin{array}{l}
		\delta g_{syn}(t)= \tilde{g}_{syn} (1- e^{-t/\tau_{\rm fast} }) \  (0<t<T), \\	
		\delta g_{syn}(t)= \tilde{g}_{syn} (1-e^{-T/\tau_{\rm fast} }) e^{-(t-T)/\tau_{\rm slow}},\  (T<t). 
	\end{array}
\end{eqnarray}
where $\tilde{g}_{syn}= g_{syn} \alpha_{syn}/ (\alpha_{syn}+ \beta_{syn})$ is the maximal synaptic conductance and $\tau_{\rm fast}= (\alpha_{syn}+\beta_{syn})^{-1}$, $\tau_{\rm slow}= \beta_{syn}^{-1}$ are the synaptic time constants. 
The conductance change is approximated by an exponential function
\begin{eqnarray}
	\begin{array}{l}
		\delta g_{syn}(t)  \approx	A_{syn} / \tau_{\rm slow} e^{- t/\tau_{\rm slow}  }, 
	\end{array}
\end{eqnarray}
where $A_{syn}= \int_0^{\infty} \delta g_{syn}(t) dt= c_{syn} \left\{  T+ (\tau_{\rm slow}- \tau_{\rm fast}) (1- e^{-T/\tau_{\rm fast} }) \right \}$. 
If the synaptic input rate $\lambda_{E, I}$ is relatively high, it is possible to apply the diffusion approximation~\cite{Lansky87,Burkitt01,Richardson05} to the excitatory and inhibitory conductances as follows, 
\begin{eqnarray}
	\beta^{-1}_{\scalebox{0.7}{AMPA} } \frac{d g_E}{dt}	&\approx& -g_E+ A_{\scalebox{0.7}{AMPA} } \left( \lambda_E+ \sqrt{\lambda_E} \xi_E(t) \right),  \\
	\beta^{-1}_{\scalebox{0.7}{GABA} } \frac{d g_I}{dt}	&\approx& -g_I+ A_{\scalebox{0.7}{GABA} } \left( \lambda_I+ \sqrt{\lambda_I} \xi_I(t) \right),  
\end{eqnarray}  
where $\xi_{E (I)}(t)$ is the Gaussian white noise. 
The excitatory (inhibitory) synaptic conductance can be decomposed into the mean and fluctuations as follows:
\begin{eqnarray} 
	g_{E (I)} = g_{E (I), 0}+ g_{E (I), F}(t), \label{eq:mean-var}
\end{eqnarray} 
where $g_{E, 0}=  A_{\scalebox{0.7}{AMPA} } \lambda_E,$ $g_{I, 0}=  A_{\scalebox{0.7}{GABA} } \lambda_I$. 	

By substituting (\ref{eq:mean-var}) into (\ref{eq:lif}) and replacing the voltage with the resting value, the voltage equation can be written as follows:
\begin{eqnarray}
	C \frac{dV}{dt} &=& -g_{\rm tot} (V-E_{\rm tot})- g_{E, F}(t) (V-E_E)- g_{I, F}(t) (V-E_I)  \nonumber \\
			&\approx & -g_{\rm tot} (V-E_{\rm tot})- g_{E, F}(t) (E_{\rm tot}-E_E)- g_{I, F}(t) (E_{\rm tot}-E_I).  \ 
\end{eqnarray}
where, $g_{\rm tot}$ and $E_{\rm tot}$ are the effective conductance and resting potentials given by
\begin{eqnarray}
	g_{\rm tot}=	g_L+ g_{E, 0}+ g_{I, 0},\ 
	E_{\rm tot}=	\left( g_L E_L + g_{I, 0} E_I \right)/ g_{\rm tot}. \nonumber
\end{eqnarray} 
If the synaptic time constants are small ($\beta_{\scalebox{0.7}{AMPA}, \scalebox{0.7}{GABA}} \gg 1$), the conductance fluctuations are approximated by white noise (white noise limit~\cite{Lindner06}), 
\begin{eqnarray}
g_{E, F}(t) \approx A_{\scalebox{0.7}{AMPA} } \sqrt{\lambda_E} \xi_E(t),\  g_{I, F}(t) \approx A_{\scalebox{0.7}{GABA} } \sqrt{\lambda_I} \xi_I(t). \nonumber
\end{eqnarray}
Accordingly, we obtain the corresponding OU model:
\begin{eqnarray}
	&&	C \frac{dV}{dt} \approx - g_{\rm tot} (V-E_{\rm tot})+ \sigma_0 \xi(t), \\
	&&	\sigma^2_0= A^2_{\scalebox{0.7}{AMPA} } E^2_{\rm tot} \lambda_E+ A^2_{\scalebox{0.7}{GABA} } (E_{\rm tot}-E_I)^2  \lambda_I. 
\end{eqnarray}

%  We would like to thank you for \textbf{following the instructions above} very closely in advance. It will definitely save us lot of time and expedite the process of your paper's publication.

% You may incorporate your references as follows in your main tex file.
% Using BibTex is not recommended but can be handled.


\begin{thebibliography}{99}

%  \bibitem{Journal-Template}
%     \newblock  FirstNameInitial.  MiddleNameInitial. LastName,     % first name middle initial. and then last name.  Only the first character in the paper title is capitalized.
%     \newblock Title of the paper,
%     \newblock \emph{Name of the Journal}, \textbf{Volume} (Year), StaringPage--EndingPage.


\bibitem{Taylor61}  
\newblock		L. R. Taylor,
\newblock		Aggregation, variance and the mean. 
\newblock		\emph{Nature} \emph{189}, (1961), 732--735.

\bibitem{Anderson89}
\newblock		R. M. Anderson and R. M. May,
\newblock		Epidemiological parameters of HIV transmission.
\newblock		\emph{Nature} \emph{333}, (1988), 514--519.

\bibitem{Kendal87}
\newblock		W. S. Kendal and P.  Frost,
\newblock		Experimental metastasis: a novel application of the variance-to-mean power function.
\newblock		\emph{J. Natl. Cancer Inst.} \emph{79}, (1987), 1113--1115.

\bibitem{Kendal04}  
\newblock		W. S. Kendal,
\newblock		A scale invariant clustering of genes on human chromosome 7. 
\newblock		\emph{BMC Evol. Biol.} \emph{4}, (2004), 3.

\bibitem{Fronczak10}
\newblock		A. Fronczak and P. Fronczak,
\newblock		Origins of Taylor's power law for fluctuation scaling in complex systems. 
\newblock		\emph{Phys. Rev. E} \emph{81}, (2010), 066112.

\bibitem{Barabasi04}
\newblock		M. A. de Menezes and A. L. Barabasi,
\newblock		OFluctuations in network dynamics. 
\newblock		\emph{Phys. Rev. Lett.} \emph{92}, (2004), 028701.

\bibitem{Eislter08}
\newblock		Z. Eisler, I. Bartos and J. Kertesz,
\newblock		Fluctuation scaling in complex systems: Taylor's law and beyond. 
\newblock		\emph{Adv. Phys.} \emph{57}, (2008), 89--142.

\bibitem{Cox66}
\newblock		D. R. Cox and P. A. W. Lewis, 
\newblock		The Statistical Analysis of Series of Events.
\newblock		London: Chapman and Hall, 1966.

\bibitem{Daley02}
\newblock		D. J. Daley and D. Vere-Jones, 
\newblock		An Introduction to the Theory of Point Processes Vol. 1.
\newblock		New York: Springer, 2002.

\bibitem{Snyder75}
\newblock		D. L. Snyder,
\newblock		Random Point Processes.
\newblock		New York: John Wiley \& Sons, Inc, 1975.

\bibitem{Johnson96}
\newblock		D. H. Johnson,
\newblock		Point process models of single-neuron discharges.
\newblock		\emph{J. Comput. Neurosci.} \emph{3}, (1996), 275--299.

\bibitem{Ogata88}
\newblock		Y. Ogata,
\newblock		Statistical models for earthquake occurrences and residual analysis for point processes. 
\newblock		\emph{J. Amer. Statist. Assoc.} \emph{83}, (1988), 9--27.

\bibitem{Bremaud81}
\newblock		P. Bremaud,
\newblock		Point Processes and Queues.
\newblock		New York: Springer, 1981.

\bibitem{Averbeck09}
\newblock		B. B. Averbeck,
\newblock		Poisson or not Poisson: differences in spike train statistics between parietal cortical areas.
\newblock		\emph{Neuron} \emph{62}, (2009), 310--311.

\bibitem{Cox62}
\newblock		D. R. Cox,
\newblock		Renewal Theory.
\newblock		London: Chapman and Hall, 1962.

\bibitem{Kampen92}
\newblock		N. G. van Kampen,
\newblock		Stochastic Processes in Physics and Chemistry 2nd Ed. 
\newblock		Amsterdam: North-Holland, 1992.

\bibitem{Wan82}
\newblock		F. Y. M. Wan and H. C. Tuckwell,
\newblock		Neuronal firing and input variability.
\newblock		\emph{J. Theoret. Neurobiol.} \emph{1}, (1982), 197--218.

\bibitem{Ditlevsen05}
\newblock		S. Ditlevsen and P. Lansky,
\newblock		Estimation of the input parameters in the Ornstein-Uhlenbeck neuronal model.
\newblock		\emph{Phys. Rev. E.} \emph{71}, (2005), 011907.

\bibitem{Tuckwell88}
\newblock		H. C. Tuckwell,
\newblock		Introduction to Theoretical Neurobiology Vol. 2. 
\newblock		New York: Cambridge University Press, 1988.

\bibitem{Nobile85}
\newblock		A. G. Nobile, L. M. Ricciardi L. and Sacerdote,
\newblock		Exponential trends of Ornstein-Uhlenbeck first-passage-time densities.
\newblock		\emph{J. Appl. Probab.} \emph{22}, (1985), 360--369.

\bibitem{Ricciardi88}
\newblock		L. M. Ricciardi L. and S. Sato,
\newblock		First-passage-time density and moments of the Ornstein-Uhlenbeck process.
\newblock		\emph{J. Appl. Probab.} \emph{25}, (1988), 43--57.

\bibitem{Siegert51}
\newblock		A. J. F. Siegert,
\newblock		On the first passage time probability problem. 
\newblock		\emph{Phys. Rev.} \emph{81}, (1951), 617--623.

\bibitem{Roy69}
\newblock		B. K. Roy and D. R. Smith,
\newblock		Analysis of the exponential decay model of the neuron showing frequency threshold effects.
\newblock		\emph{Bull. Math. Biophys.} \emph{31}, (1969), 341--357.

\bibitem{Abramowitz65}
\newblock		M. Abramowitz and I. Stegun,
\newblock		Handbook of Mathematical Functions.
\newblock		New York: Dover, 1965.

\bibitem{Keilson75}
\newblock J. Keilson and H. F. Ross,
\newblock  Passage time distribution for Gaussian Markov (Ornstein-Uhlenbeck) statistical processes, 
\newblock  in \emph{Selected Tables in Mathematical Statistics} \emph{3},
                American Mathematical Society, (1975), 233--328.

\bibitem{Inoue95}
\newblock		J. Inoue, S. Sato and L. M. Ricciardi,
\newblock		On the parameter estimation for diffusion models of single neuron's activities. I. Application to spontaneous activities of mesencephalic reticular formation cells in sleep and waking states.
\newblock		\emph{Biol. Cybern.} \emph{73}, (1995), 209--221.

\bibitem{Vilela09}
\newblock		R. D. Vilela  and B. Lindner,
\newblock		Are the input parameters of white noise driven integrate and fire neurons uniquely determined by rate and {CV}?
\newblock		\emph{J. Theor. Biol} \emph{257}, (2009), 90--99.

% neuron model

\bibitem{Destexhe1998} 
\newblock		A. Destexhe, Z. Mainen, and  T. J. Sejnowski,  
\newblock		Kinetic models of synaptic transmission.
\newblock		in \emph{Methods in Neuronal Modeling} (pp.1--26) (eds. C. Koch and I. Segev), Cambridge, MA: MIT Press (1998), 1--26. 

%  diffusion approximation
\bibitem{Lansky87}
\newblock		P. Lansky, and V. Lanska, 
\newblock		Diffusion approximation of the neuronal model with synaptic reversal potentials. 
\newblock		\emph{Biol. Cybern.}, \emph{56}, (1987), 19--26.

\bibitem{Burkitt01}
\newblock		A. N. Burkitt,  
\newblock		Balanced neurons: analysis of leaky integrate-and-fire neurons with reversal potentials. 
\newblock		\emph{Biol. Cybern}, \emph{85}, (2001), 247--255.

\bibitem{Richardson05}
\newblock		M. J. Richardson, and W. Gerstner,  
\newblock		Synaptic shot noise and conductance fluctuations affect the membrane voltage with equal significance. 
\newblock		\emph{Neural Comput.}, \emph{17}, (2005), 923--947.

\bibitem{Lindner06}
\newblock		B. Lindner, and A. Longtin, 
\newblock		Comment on ``Characterization of Subthreshold Voltage Fluctuations in Neuronal Membranes," by M. Rudolph and A. Destexhe. 
\newblock		\emph{Neural Comput.}, \emph{18}, (2006), 1896--1931.

%  section 4
\bibitem{ChurchlandMM10}
\newblock		M. M. Churchland et al.,
\newblock		Stimulus onset quenches neural variability: a widespread cortical phenomenon.
\newblock		\emph{Nat. Neurosci.} \emph{13}, (2010), 369--378.

\bibitem{ChurchlandAK11}
\newblock		A. K. Churchland et al.,
\newblock		Variance as a signature of neural computations during decision making. 
\newblock		\emph{Neuron} \emph{69}, (2011), 818--831.

\bibitem{ChurchlandMM12}
\newblock		M. M. Churchland and L. F. Abbott,
\newblock		Two layers of neural variability.
\newblock		\emph{Nat. Neurosci.} \emph{15}, (2012), 1472--1474.

% discussion section
\bibitem{Burkitt06}
\newblock		A. N. Burkitt,
\newblock		A review of the integrate-and-fire neuron model: I: homogeneous synaptic input.
\newblock		\emph{Biol. Cybern.} \emph{95}, (2006), 1--19.

\bibitem{Troy92}
\newblock		J. B. Troy and J. G. Robson,
\newblock		Steady discharges of X and Y retinal ganglion cells of cat under photopic illuminance.
\newblock		\emph{Vis. Neurosci.} \emph{9}, (1992), 535--553.

\bibitem{Tolhurst81}
\newblock		D. J. Tolhurst, J. A. Movshon and I. D. Thompson,
\newblock		The dependence of response amplitude and variance of cat visual cortical neurones on stimulus contrast.
\newblock		\emph{Exp. Brain Res.} \emph{41}, (1981), 414--419.

\bibitem{Shadlen98}
\newblock		N. M. Shadlen and W. T. Newsome,
\newblock		The variable discharge of cortical neurons: implications for connectivity, computation, and information coding.
\newblock		\emph{J. Neurosci.} \emph{18}, (1998), 3870--3896.

\bibitem{Lerchner06}
\newblock		A. Lerchner et al.,
\newblock		Response variability in balanced cortical networks
\newblock		\emph{Neural Comput.}, \emph{18}, (2006), 634--659.
             
\end{thebibliography}
\end{document}